# From Protostars to Planetary Systems : FUV Spectroscopy of YSOs, Protoplanetary Disks, and Extrasolar Giant Planets

Paul Scowen

School of Earth & Space Exploration
Arizona State University
PO Box 871404, Tempe, AZ 85287-1404

(480) 965-0938

paul.scowen@asu.edu

Rolf Jansen (Arizona State University)

Matthew Beasley (University of Colorado – Boulder)

Steve Desch (Arizona State University)

Alex Fullerton (STScI)

Mark McCaughrean (University of Exeter)

Sally Oey (University of Michigan)

Debbie Padgett (IPAC / Caltech)

Aki Roberge (NASA – GSFC)

Nathan Smith (University of California – Berkeley)

*Protostars to Planets***Abstract**

The last two decades have seen remarkable progress in our long-standing goal of determining the abundance and diversity of worlds in the Galaxy. Understanding of this subject involves tracing the path of interstellar material from dense cloud cores, to young stellar objects, protoplanetary disks, and finally extrasolar planets. Here we discuss the critical information provided on these objects by point-source far-ultraviolet spectroscopy with a large aperture, high resolution spectrograph of a large sample of unique protostellar and protoplanetary objects that will leverage our existing knowledge to lay out a path to new and powerful insight into the formation process. We lay out a systematic case of coordinated observations that will yield new knowledge about the process of assembly for both protostellar and protoplanetary systems – that addresses specific uncertainties in our current knowledge and takes advantage of potential new technologies to acquire the data needed.**Introduction & Scientific Context**

The collapse of an interstellar (IS) molecular cloud core naturally leads to the formation of a disk of gas and dust around the growing protostar. Further growth of the protostar occurs by accretion of material through the disk. Planets eventually form in this disk, although it was generally thought that this did not occur until accretion onto the star had largely ceased and the disk was "passive". However, it is becoming clear that the late stages of star formation and the early stages of planet formation can occur simultaneously, and we need to know more details about the former process to understand how it may affect the latter. The addition of NUV and FUV spectroscopic data to this problem allows the use of line flux measurements to study material accreting onto young stars from their accretion disks, as it shocks and is heated to about $10^6$ K before being radiatively cooled. Such infall is not sufficient to produce X-rays, but the first observable emission from such shocks is from gas at nearly $10^5$ K using lines such as C IV, Si IV, etc. This emission represents the least processed radiation produced by accretion processes and provides direct access to accurate estimates of accretion rates, filling factors, etc. In addition to hot emission lines, the FUV can be used to probe molecular hydrogen in disks and the near circumstellar environment around T-Tauri stars (TTS). Current estimates of the assembly time for Jupiter mass planets does not match the measured lifetimes of accretion disks from thermal dust emission, so it is expected that gas survival rates are longer. $H_2$ is widely expected to be the primary gas constituent of circumstellar disks, so direct measurements of molecular hydrogen will solve the disagreement. Re-analysis of the low resolution *IUE* final archive spectra has revealed FUV fluorescent $H_2$ emission from a dozen classical TTS. In similar fashion, a large aperture telescope combined with a next generation high throughput, high resolution FUV spectrograph would be uniquely suited to this task, while affording a considerable increase in collecting area, throughput and DQE.

**Compelling Science Themes Based on Recent Advances**

**FUV Emission Lines as Probes of Mass Accretion in Young Stars**

A variety of observations of young stars, including emission line profiles, spectral energy distributions, and variability, indicate that stellar magnetic fields channel material from

*Scowen et al.*     1

circumstellar (CS) accretion disks onto their stars. The CS disk is truncated close to the star by the stellar magnetic field, and disk material captured by this field produces a strong shock as it slides down the gravitational potential of the star and strikes the photosphere. The accretion process also creates highly supersonic bipolar outflows that lie perpendicular to the disk planes, and shock waves in these flows generate chains of emission line knots known as Herbig-Haro objects.

The far-ultraviolet (FUV) region of the spectrum is particularly important for testing models of accretion in young stars because it allows observation of the $10^5 - 10^6$ K gas produced immediately behind the accretion shocks. As the shocked gas cools, it emits strongly in FUV lines, which are optically thin. Young low mass stars are also magnetically active, so there is typically a contribution to the FUV lines from the transition regions and chromospheres. Indeed, for X-ray emission, magnetic coronal emission in these stars dominates over the emission produced in the accretion shocks. However, in classical (i.e. accreting) T-Tauri stars (CTTS), the accretion shock produces the bulk of the FUV line emission, since it is enhanced by at least an order of magnitude over the level expected from magnetic activity and the FUV line luminosity correlates strongly with the mass accretion rate. The observed FUV line luminosities are also consistent with the predictions of shock models with accretion rates typical of CTTS.

Because the FUV lines of CTTS form primarily in the accretion shocks, they open up new ways to study the physics of magnetospheric accretion. Understanding this process is essential if we are to develop an accurate picture of how planetary systems form. Accretion and outflow are the means by which young stars and their disks redistribute mass and angular momentum, and these processes also control heating in the disk. The next two sections outline several ways in which the unique new FUV spectroscopic capability we have described will address fundamental unsolved questions concerning magnetospheric accretion onto young stars.

**Accretion Rates in Herbig Ae/Be Stars**

While the existence of magnetospheric accretion in CTTS has much observational support, the situation for the Herbig Ae/Be stars, which are higher mass analogs of T-Tauri stars, is much less clear. Our understanding of the close CS environment in these higher mass stars is very incomplete, and the role that accretion plays in the observed phenomena is largely unknown. CS disks have been imaged around a few Herbig Ae/Be stars; however, their frequency has not been determined.

Accretion onto low mass young stellar objects (YSOs) produces hot photospheric material as a result of reprocessing the radiation from the accretion shock. The heated photosphere gives rise to both optical and UV excess continuum emission which appears to fill in or "veil" the stellar photospheric absorption lines. This is relatively easy to see in CTTS spectra, since the material producing the veiling is much hotter than the stellar photosphere. However, for Herbig Ae/Be stars, it is difficult to distinguish the excess continuum emission from the photospheric continuum, since the temperatures of the hot spots and the stellar photospheres are comparable (T ~ $10^4$ K). In addition, Herbig Ae/Be stars usually display evidence for strong winds in high resolution optical line profiles and in the profiles of near-UV Mg II lines. As a result, it is extremely difficult to use optical or near-UV lines as accurate tracers of disk accretion for these objects.

Fortunately, the hotter FUV emission lines formed in the accretion shock provide excellent probes of mass accretion in Herbig Ae/Be stars, and in fact will be the *only* way to accurately measure accretion rates in many cases. An atlas of Herbig Ae/Be stars observed with *IUE* shows that many of them do show FUV emission lines such as C IV at 1550Å. It has been shown that for a sample of stars with masses intermediate between typical CTTS and Herbig Ae/Be stars, the FUV line strengths correlate well with the mass accretion rates, lending further support to their use as rate tracers in Herbig Ae/Be stars.

Our survey will be ideally suited to study accretion tracer lines such as O VI and C III in a large sample of Herbig Ae/Be stars, and will allow us to explore how much accretion affects these higher mass pre-main sequence stars. Our study will address several key questions: Are the accretion rates for Herbig Ae/Be stars larger on average than for CTTS? Current studies of intermediate mass YSOs find a correlation between stellar mass and accretion rate, so we might expect the Herbig Ae/Be stars to show large mass accretion rates as a whole. But only a large sample of such objects observed at various ages can establish how these systems form. Is the accretion on Herbig Ae/Be stars controlled by stellar magnetic fields? Conventional wisdom assumes the picture we have for CTTS applies to the higher mass stars as well, but the origin of magnetic fields in these fully radiative stars is not understood. If stellar magnetic fields do play a role, we expect rotational modulation the FUV line profiles, as described in the next subsection. Is the typical disk accretion timescale in Herbig Ae/Be stars equal to or shorter than that of CTTS? If they really do have large accretion rates as a whole, the disk lifetimes may be shorter in these stars, which would have broad implications for planet formation in general around higher mass stars, and gas giant formation in particular.

**Mapping the Shock Waves and Accretion Columns in CTTS**

The FUV emission lines accessible in our survey provide a unique means to study both the physics of the accretion shock and the basic geometry of magnetospheric accretion in young stars. Current models assume the magnetic field serves only to direct the accretion flow, but that the shock itself is not influenced by the field. However, it is well known that transverse magnetic fields can soften shocks in the ISM. Observations of infall signatures in optical spectral lines in CTTS constrain the pre-shock velocity, while the FUV line ratios strongly constrain the velocity (through the peak temperature reached in the shock) and the density of the shocked material. Hence, the only way to explore the degree to which magnetic fields influence accretion shocks on the surface of CTTS is with a multiwavelength approach that involves both FUV and optical line diagnostics.

A way in which our survey can make a huge impact on the field of star formation is by mapping the actual geometry of magnetospheric accretion. Our program will be the first to connect accretion shocks on the stellar photosphere (traced by FUV lines), with accretion columns in the extended magnetosphere and with outflows generated from the inner disk surface, both of which emit optical line emission. The first magnetospheric accretion theories assumed that the dipole axis of the magnetic field is aligned with the stellar rotation axis, so that the accretion occurs in an axisymmetric ring near the rotation poles. However, observations of rotational modulation of accretion signatures show that the geometry is more complex, and a tilted dipole geometry is popular now, with the accretion still occurring near the rotation poles. Observational support for this picture is found in time series analysis of Balmer line profile variations in the CTTS SU Aur - accretion and wind signatures in the Hα and Hβ line profiles are rotationally modulated and vary exactly 180º out of phase with one another, a natural prediction of a tilted dipole geometry.

However, while similar optical line profile variability has been observed in one other CTTS, systematic variability is not generally seen. Part of the difficulty certainly lies in the fact that variations in the accretion rate also affect the Balmer line profiles. In addition, the Balmer lines form over an extended zone including the magnetosphere and the wind/jet launching region in these stars. Hence, it is difficult to disentangle where various features in the profiles form and reconstruct the geometry of the line formation region. In addition, spectropolarimetric observations of CTTS show that the magnetic fields are not simply dipolar. As a result, the magnetic field geometry controlling accretion in CTTS as a whole is not well understood. This lack of knowledge obviously impacts accretion studies of these stars, but may also affect our understanding of the origin of winds and jets observed from YSOs. For example, one of the two leading theories for the origin of the winds and jets in these stars relies on the interaction of the magnetic field with the inner disk (e.g. the "X-wind" theory) and is critically dependent on the geometry of this interaction.

**The Evolution of Gas in Protoplanetary Disks**

Ideally, a CS disk evolves from a massive gas-rich remnant of star formation (a *primordial disk*) to a relatively low mass, gas-free planetary system. An intermediate stage is the *debris phase*, during which the disk is composed of material produced by the destruction of planetesimals. Debris disks are basically young, dense versions of our asteroid and Kuiper belts. However, the details of disk evolution are not known, including the times and conditions required for formation of various planetary bodies. In the traditional picture, giant planets form by accretion of gas onto a massive solid core, which must form before the primordial molecular gas in the disk dissipates. Observations suggest that the typical gas dissipation timescale is shorter than about 10 Myr, possibly as short as 1 Myr in regions of high-mass star formation. Massive core formation takes at least a few Myr for solar-type stars, and much longer for low-mass stars. This has led some workers to suggest alternative theories for giant planet formation, i.e. that they form very rapidly by direct collapse in gravitationally unstable disks. However, it is not yet absolutely clear that core-accretion models must be rejected, due to the uncertain timescale for primordial gas dissipation around different types of stars.

The goal of this research is to characterize the gas in protoplanetary disks, as it evolves from primordial molecular gas to secondary gas produced by evaporation of planetesimals. This will provide critical constraints on
1) the process by which giant planets form,
2) the dynamics of planetary bodies and dust in young CS disks, and
3) the nature and history of volatile materials in terrestrial planets.

**Primordial Gas and Giant Planet Formation**

The decisive test of core-accretion theory may come from studies of young low-mass stars. Giant planets are found around M-stars, but are extremely hard to form by core-accretion unless the primordial gas lifetime is longer than around solar-type stars. Therefore, comparison between the frequency of planets around low-mass stars and the typical $H_2$ lifetime in their disks will be a powerful constraint on giant planet formation theory.

Difficulties in measuring disk gas masses are responsible for the current uncertainty about the primordial gas lifetime. The most commonly used molecular gas tracer, carbon monoxide, can freeze out onto dust grains in the cold outer regions and midplanes of disks, and is not a reliable proxy for total gas mass. Direct observations of the most abundant molecular gas, $H_2$, are necessary to determine total disk gas masses. The FUV electronic transitions of $H_2$ are very strong and provide access to gas with a wide range of temperatures, from a few Kelvins to thousands of Kelvins. This can be through observation of line of sight $H_2$ absorption or observation of fluorescent $H_2$ emission produced by UV pumping. Such studies have shown a surprising diversity of $H_2$ gas characteristics in systems with ages in the critical range from ~ 1 – 10 Myr.

While the existing FUV spectra of protoplanetary disks are intriguing, only a few systems have been studied, and it is not known if they are typical. The most critical shortcoming of the sample is that it is heavily weighted toward the FUV-bright A-stars and currently contains only one M-star (the relatively old system AU Mic). *FUSE* was not sensitive enough to observe more than 3 or 4 of the nearest young M-stars. Study of the typical primordial gas lifetime in the disks of low-mass stars will require the greater sensitivity of our survey.

**Secondary Gas in Debris Disks**

After primordial gas has dissipated, secondary CS gas is produced by the destruction of planetesimals. For example, CO observed in the β Pic debris disk is definitely secondary gas, likely produced by evaporation of icy comet-like material. Secondary gas is of interest for two primary reasons. First, it is believed that most of the volatile content of terrestrial planets is delivered during the debris phase by the impact of water-rich planetesimals. The composition of planetesimals in debris disks provides insights on terrestrial planet atmospheres, including Earth as well as those to be studied by *Terrestrial Planet Finder*. Extrasolar planetesimals are far too small and faint to perform spectroscopy on individual objects, but the secondary gas in debris disks should reflect their bulk composition.

Second, dust structures seen in coronagraphic images of debris disks, like clumps and rings, have been attributed to the gravitational effect of unseen giant planets. As such, they may provide indirect evidence of young planets that cannot be detected with standard radial velocity techniques because of the intrinsic variability of young stars. Unfortunately, dynamical models of these structures do not take into account secondary gas, and rarely include gas drag at all, due to the lack of useful observational constraints. However, even *very* low gas abundances will have a dramatic effect on small grains, which are the ones most easily seen in disk images. Interpretation of dust structures will never be secure until secondary gas abundances in debris disks are measured so that gas drag may be accurately included in dynamical models.

To date, only two observational techniques have been able to firmly detect gas in debris disks: 1) UV/optical absorption spectroscopy of edge-on disks and 2) spatially-resolved optical spectroscopy of resonantly scattered emission from gaseous species. FUV spectroscopy is particularly important, since the transitions out of the ground states of the most abundant atomic species lie in the FUV (H I, C I, C II, N I, O I, etc.). Transitions of potentially abundant gaseous molecular species such as CO, OH, and $H_2O$ also lie in the FUV. Further progress on characterization of gas in debris disks will require high-resolution FUV absorption spectroscopy.

**The Composition and Structure of Extrasolar Giant Planets**

The discovery of giant planets orbiting very close to their parent stars (within 1 AU) was one of the biggest surprises of the last 10 years. It is extremely unlikely that these so-called "hot Jupiters" formed where they are currently seen. Rather, it is believed that they formed at Jupiter-like distances from their parent stars and then moved into tight orbits through the newly suggested and poorly understood process of planetary migration. If the inclination of a planet's orbit with respect to plane of the sky is high, then it may pass in front of its parent star, i.e. transit the star. Their proximity to their parent stars makes it more likely that hot Jupiters will transit. Currently, there are 5 known transiting hot Jupiters, the best-studied of which is HD 209458. Broad-band photometric light curves of HD 209458 taken during transit gave the size of the planet, allowing calculation of its mean density and giving the first real confirmation that it was indeed a gas giant.

During a transit, light from the star passes through the planet's atmosphere. This raises the exciting possibility of probing the structure and composition of extrasolar giant planets using transit spectroscopy. The first detection of an extrasolar planet atmosphere was with space-based optical absorption spectroscopy of Na I in HD 209458. The planet's atmosphere has also been detected with transit spectroscopy at Lyman-$\alpha$. This observation showed that the planet is surrounded by an extended H I atmosphere that overflows its Roche lobe: the planet is being evaporated by its central star.

These observations represent the beginning of general characterization of extrasolar giant planets. Transit spectroscopy, particularly in the FUV where the transitions of the most abundant atomic species lie, will allow us to answer many questions. What are the compositions of the atmospheres of hot Jupiters? Are they actually similar to gas giants in the Solar System (Jupiter, Saturn)? What is the structure of their outer atmospheres? How fast do they evaporate and what do they look like after significant evaporation? There are many ground-based surveys in progress searching for transiting giant planets, which seem to be beginning to bear fruit. In addition, the Discovery-class *Kepler* space mission is scheduled to be launched in 2009. The goal of this mission is the indirect detection of terrestrial planets through high-precision photometric monitoring of 100,000 stars for transits. Many transiting giant planets will be detected as well, providing an excellent set of targets for transit spectroscopy with our survey.

**Key Advances in Observation Needed**

To achieve the science goals of this program a variety of capabilities need to be implemented. The majority of the objects being targeted require the aperture of a medium to large (2-4m) UV/optical space telescope combined with a high throughput, high resolution FUV spectrograph that is sensitive down to 100nm to provide access to key diagnostic lines. Over the next decade the specific technological capabilities that need to be developed include the design of next generation FUV reflective coatings and next generation FUV detectors that can yield remarkable advances in detection efficiency for a minimal investment.

**Four Central Questions to be Addressed**

1. What are the preferred modes of accretion onto young low-mass stars? What do these modes tell us about the earliest of stages of formation of planetary systems around this size of star?
2. Similarly, what are the preferred modes of accretion onto higher mass stars such as Herbig Ae/Be stars, and what do differences in these modes tell us about the process of planet building? In particular what does this tell us about the mechanism that forms giant planets?
3. What are the gas dissipation timescales in early protoplanetary disks and what restrictions does this place on the formation of giant planets versus terrestrial planets in terms of cutting off the supply of accreting material?
4. What is the mechanism by which volatiles are delivered to terrestrial planets and thereby create the atmospheres and produce water-rich cnvironments on what could become physical analogs to our own Earth?

**Area of Unusual Discovery Potential for the Next Decade**

While the science program in this paper have defined a loose set of specifications (see Table 1), it should be recognized that the opportunity for truly unique discovery is made possible by the combination of two new technological developments over the next decade: new low-maintenance, low-risk, high reflectivity FUV optical coatings, and next generation multi-channel plate detectors that yield high DQEs in large enough formats to allow simple construction of focal plane arrays with a small enough resolution element pitch to support high resolution spectroscopy.

| Parameter | Specification | Justification |
|---|---|---|
| Aperture | 2-4m | This is driven by the limiting magnitudes needed traded against the necessary exposure times to achieve them – the larger the better |
| Stability | A small percentage of a pixel | To allow the stable photometry and astrometric measurements necessary to achieve the science goals |
| Photometric Stability | Combination of gain, A/D conversion and QE need to be stable to better than $10^{-5}$ | Again to provide the photometric stability to achieve the science goals of the project |
| Coatings | Development of stable, high-reflectivity FUV mirror coatings | To provide high throughput access to the FUV (below 115nm) while minimizing risk to the optical reflectivity of an optical system |
| FUV Detectors | Development of next generation MCP technology | To provide a low-cost, high QE, robust solution to allow efficient observations of FUV emission, below 115nm to as low as 100nm |
| FUV Spectroscopic Resolution | R > 30,000 | To enable sufficient resolution to see structure and dynamics of emission from science targets |

**Table 1**: Science Driven General Specifications